\documentstyle[12pt]{article}

\newcommand{\vp}{\varphi}
\newcommand{\eps}{\epsilon}
\newcommand{\epsb}{\bar \epsilon}
\newcommand{\Z}{{\sf Z \!\!\! Z}}
\newcommand{\hsp}{\hspace*{-7mm}}

\makeatletter
\@addtoreset{equation}{section}
\makeatother

\title{Exact Supersymmetry on the Lattice}

\author{W. Bietenholz\\
HLRZ c/o Forschungszentrum J\"{u}lich\\
D-52425 J\"{u}lich, Germany \\ \\
Preprint HLRZ 1998-24}

\begin{document}
\maketitle
\begin{abstract} \normalsize






{\it We discuss the possibility of representing supersymmetry
exactly in a lattice discretized system. In particular,
we construct a perfect supersymmetric action 
for the Wess-Zumino model.}

\end{abstract}

\section{Introduction}

Lattice simulations of supersymmetric systems usually apply
formulations, which reveal the supersymmetry 
in the continuum limit but not in the lattice discretized 
version \cite{Montvay}. This note
addresses the issue of a lattice action, which directly
displays a continuous form of supersymmetry.
Some related works are listed in Ref. \cite{susylat}.

\subsection{A simple supersymmetric model}

We first consider a simple 2d SUSY model \cite{GS}
given in the continuous Euclidean space by the Lagrangian
\begin{equation}
{\cal L} = \bar \psi \gamma_{\mu} \partial_{\mu} \psi
+ \partial_{\mu} \vp \partial_{\mu} \vp \ ,
\end{equation}
where $\psi$ is a real ``Majorana spinor'' 
\footnote{Strictly speaking, there are no Majorana spinors in
Euclidean space, we just refer to the Euclidean version of the
corresponding formulae in Minkowski space. A definition is
given for instance in Ref. \cite{Gol}, based on a Euclidean
analog of charge conjugation.
Note that $\bar \psi$ and $\psi$ are not independent.}
and $\vp$ is a real scalar field.
Many qualitative  features with respect to a lattice formulation
are the same as in the Wess-Zumino model.
The action $S$ is invariant under the field transformations
\begin{equation} \label{trafo}
\delta \psi = - \gamma_{\mu} \partial_{\mu} \vp \ \eps
\quad , \quad \delta \vp = \epsb \ \psi \ ,
\end{equation}
where the two components of the transformation parameter
$\ \eps \ $ are real Grassmann variables.
As an important property, we note that the supersymmetric
generator forms a closed algebra with the translation operator,
\begin{equation} \label{transla}
[\delta_{1}, \delta_{2}]\vp = ( \bar \eps_{1} \ \gamma_{\mu} \
\eps_{2} \ - \ \bar \eps_{2} \ \gamma_{\mu} \ \eps_{1} ) \
\partial_{\mu} \vp \ .
\end{equation}

\subsection{Ansatz for a lattice formulation} \label{ansatz}

Let us consider a rather general ansatz for a lattice discretization
of this free system in momentum space,
\begin{eqnarray}
S &=& \int_{B} \frac{d^{2}p}{(2\pi )^{2}} \ \Big\{
\bar \Psi (-p) [ \gamma_{\mu} \rho_{\mu}(p) + \lambda (p)]
\Psi (p) + \Phi (-p) \Omega (p) \Phi (p) \Big\} \ , \nonumber \\
\delta \Psi (p) &=& - [ \gamma_{\mu} R_{\mu}(p) + L(p)]
\Phi (p) \eps \ , \nonumber \\ \label{act}
\delta \Phi (p) &=& \epsb [u(p)+ \gamma_{\mu}v_{\mu}(p)] \Psi \ .
\end{eqnarray}
Here, $\Psi$ and $\Phi$  are the massless lattice fermion resp. scalar
field, and the new quantities, which we introduce as an ansatz for the
inverse propagators and for the transformation terms ($\rho_{\mu},
\lambda , \Omega ,R_{\mu},L,u,v_{\mu}$) are real in coordinate
space. It is desirable that they are all local, i.e. analytic in
momentum space. We require
the low energy expansion of the action to reproduce the correct
continuum limit, and the inverse propagators obey in coordinate
space the normalization conditions
\footnote{In the first expression, there is no sum over $\mu$.}
\begin{equation} \label{norm}
\sum_{x} x_{\mu} \rho_{\mu}(x) = 1 \ , \ \sum_{x} \lambda (x) = 0
\ , \ \sum_{x} x^{2} \Omega (x) = -4 \ .
\end{equation}

We do not require the lattice transformation terms to correspond 
exactly to the continuum transformations (\ref{trafo}).
Hence this ansatz includes a possible ``remnant supersymmetry'' of the
lattice action, similar to the Ginsparg-Wilson relation for
the chiral symmetry \cite{ML}.
The general (remnant) supersymmetry requirement $\delta S =0$
amounts to
\begin{eqnarray} \hspace{-13mm} &&
-R_{\mu}(-p) [\rho_{\mu}(p)-\rho_{\mu}(-p)]+L(-p)[\lambda (-p)-
\lambda (p) ] 
+ u(p) [\Omega (p) + \Omega (-p)] = 0 \ , \nonumber \\ \hspace{-10mm}&&
R_{\mu}(-p)[\lambda (-p)-\lambda (p)] + L(-p) [ \rho_{\mu}(-p) - 
\rho_{\mu}(p)] 
+ v_{\mu}(p) [\Omega (p) + \Omega (-p)] = 0 \ .
\end{eqnarray}
It is sensible to assume the following symmetry properties in the
action: $\rho_{\mu}$ is odd in the $\mu$ component and even in all 
other directions, while the Dirac scalars
$\lambda$ and $\Omega$ are entirely even. Then the conditions simplify to
\begin{eqnarray}
-R_{\mu}(-p) \rho_{\mu}(p) + u(p) \Omega (p) &=& 0 \ ,\nonumber \\
- L(-p) \rho_{\mu}(p) + v_{\mu}(p) \Omega (p) &=& 0 \ .
\end{eqnarray}
Remarkably, $\lambda $ does not occur any more in these conditions.

Finally, we assume for the transformation terms $R_{\mu}$ and $L$
the same symmetries as for $\rho_{\mu},\lambda$, respectively,
which leads to
\begin{eqnarray}
-R_{\mu}(p) \rho_{\mu}(p) &=& u(p) \Omega (p) \ ,
\nonumber \\
L (p) \rho_{\mu}(p) &=& v_{\mu}(p) \Omega(p)
\qquad \quad (\mu=1,2) . \label{condgen}
\end{eqnarray}

The translation operator is identified from
\begin{eqnarray} \hspace{-8mm}
[\delta_{1},\delta_{2}] \Phi_{x} &=& \sum_{y} [ \bar \epsilon_{1} \ 
Q(x-y) \ 
\epsilon_{2} \ - \ \bar \epsilon_{2} \ Q(x-y) \ \epsilon_{1} ] 
\Phi_{y} \ , \nonumber \\ \hspace{-8mm}
Q(x-y) &=& \sum_{z}[u(x-z) + \gamma_{\nu} v_{\nu}(x-z)] \
[\gamma_{\mu} R_{\mu}(z-y) +L(z-y)] \ ,
\end{eqnarray}
but this general form is not immediately instructive.

Let us consider simple solutions for the case 
$u=1$, $v_{\mu}=L=0$.
\footnote{Note that the two transformation terms with an unusual Dirac
structure, $L$ and $v_{\mu}$, can only come into play simultaneously.}
The standard lattice action,
$\rho_{\mu}(p) = i \bar p_{\mu} := i \sin p_{\mu}$,
$\Omega (p) = \hat p^{2} := \sum_{\mu} [2 \sin (p_{\mu}/2)]^{2}$,
requires
\begin{displaymath}
R_{\mu}(p) = \frac{\hat p^{2}}{\bar p^{2}} \ i \bar p_{\mu} \ ,
\end{displaymath}
which is singular at $p =(\pi ,0)$, $(0,\pi)$ and $(\pi ,\pi)$,
hence the transformation is non-local.
An obvious concept to simplify $R_{\mu}$ 
-- and to obtain the same dispersion relation for fermion and scalar --
is the use of the
same lattice differential operator for the scalar and the fermion part
of the action. 

One way to do so is to set $\Omega(p)
= \bar p^{2}$, $R_{\mu}(p)=\rho_{\mu}(p)=i\bar p_{\mu}$, which is local but
affected by 
doubling, both, for the fermion as well as the scalar.
In the present model, unlike the case of
the Wess-Zumino model \cite{Jap},
we cannot treat them by adding Wilson terms ($(r/2) \cdot\hat p^{2}$),
because terms of this kind alter $\Omega$ but not $\rho_{\mu}$
(the fermionic Wilson terms contributes to $\lambda$).
Hence $R_{\mu}$ gets complicated and non-local again,
$R_{\mu}(p) = i \bar p_{\mu}(1 + (r/2)\hat p^{2}/\bar p^{2})$.

If one is ready to accept non-locality, then it looks simpler to
adjust the differential operators the other way round,
$\rho_{\mu}(p)=R_{\mu}(p)= i \hat p_{\mu}$,
as suggested in Ref. \cite{China}, and $\Omega (p) = \hat p^{2}$.
Then the translation operator resulting from the lattice version
of eq. (\ref{transla}) corresponds to a half-lattice shift,
whereas it is a full lattice shift for the option mentioned before.
However, the fermionic inverse propagator performs a finite gap
at the edge of the Brillouin zone, so we are dealing with a
non-locality similar to the SLAC fermions. This suggests that also this
approach fails to recover Lorentz invariance in the presence of a
gauge interaction, as was pointed out for the SLAC fermions 
on the one loop level \cite{KS}.

One can construct a number of solutions of this type by hand.
For instance, the standard action together with $u(p)= \prod_{\mu}
\cos (p_{\mu}/2)$ even provides locality, but such hand-made
constructions look hardly satisfactory. Similar
to the problem of chiral fermions on the lattice, they do not
appear promising for the consistent incorporation of interactions.
Hence we are going to follow a different, more systematic strategy.

\section{A perfect supersymmetric lattice action} \label{perfect}

Since we are considering a free theory here, we can construct
a perfect lattice action by ``blocking from the continuum'',
which corresponds to a block variable renormalization group 
transformation (RGT) with blocking factor infinity,
\begin{equation} \label{RGT}
e^{-S[ \Psi ,\Phi ]} = \int D \psi D \vp \
e^{-s[ \psi , \vp ] - T [ \Psi , \psi , \Phi , \vp ] } \ ,
\end{equation}
where $S$ is the perfect lattice action (i.e. an action without
lattice artifacts), $s$ the continuum action, 
and $T$ the transformation term.
We choose the latter such that the functional integral remains 
Gaussian,
\begin{eqnarray}
T &=& \sum_{x,y} [ \bar \Psi_{x} - \int_{C_{x}} \bar \psi (u) du ]
\ (\alpha^{f})^{-1}_{xy} \ [ \Psi_{y} - \int_{C_{y}} \psi (u) du ] 
\nonumber \\
&+& \sum_{x,y} [ \Phi_{x} - \int_{C_{x}} \vp (u) du ] \label{TT}
\ (\alpha^{s})^{-1}_{xy} \ [ \Phi_{y} - \int_{C_{y}} \vp (u) du ] \ ,
\end{eqnarray}
where $C_{x}$ is the unit square with center $x$, and $\alpha^{f}$,
$\alpha^{s}$ are arbitrary RGT parameters ($\alpha^{s}$ has to be
positive). 
The resulting perfect action reads
\begin{eqnarray} \hsp \hspace{-5mm} &&
S[\Psi ,\Phi ] = \frac{1}{(2\pi )^{2}} \int_{-\pi}^{\pi}
d^{2}p \ \Big\{ \bar \Psi (-p) \Delta^{f}(p)^{-1} \Psi (p)
+ \Phi (-p) \Delta^{s}(p)^{-1} \Phi (p) \Big\} \nonumber \\
\hsp \hspace{-5mm} && 
\Delta^{f}(p) = \sum_{l \in \Z^{2}} \frac{\Pi (p+2\pi l)^{2}}
{i (p_{\mu} + 2\pi l_{\mu}) \gamma_{\mu}} + \alpha^{f}(p) \ , \quad
\Delta^{s}(p) = \sum_{l \in \Z^{2}} \frac{\Pi (p+2\pi l)^{2}}
{(p + 2\pi l)^{2}} + \alpha^{s}(p) \ , \label{perfact}
\end{eqnarray}
where $\Pi (p) := \prod_{\mu} \hat p_{\mu}/p_{\mu}$.
Locality requires $\alpha^{f}\neq 0$, which naively breaks the chiral
symmetry. However, the latter is still present in the observables
\cite{Schwing}, and a continuous remnant form of it even persists
in the lattice action, if $\alpha^{f}(p)$ is analytic \cite{ML}.
\footnote{In this case, the full chiral symmetry of the fermion propagator
is broken only locally. This is the property denoted as Ginsparg-Wilson
relation.}

Now we consider the SUSY transformation. The variation of the 
continuum fields is given in eq. (\ref{trafo}). If we transform 
simultaneously the lattice fields as
\begin{equation} \label{ptrafo}
\delta \Psi_{x} = - \gamma_{\mu} \int_{C_{x}} \partial_{\mu} \vp (u)
du \ \eps \ , \quad
\delta \Phi_{x} = \epsb \int_{C_{x}} \psi (u) du \ ,
\end{equation}
then all the square brackets in the expression for $T$ (eq. (\ref{TT}))
remain invariant -- and so does the continuum action -- hence
$\delta S =0$.

Everything is consistent since we block the fields as well as
their variations from the continuum.
Note that this is not a solution along the lines of section \ref{ansatz}, 
because the transformations of the
lattice fields are not expressed directly in terms of lattice fields.
(The special case of a $\delta$ function RGT, $\alpha^{s},
\ \alpha^{f} \to 0$, is an exception, see below.)
Hence the solution is somehow implicit.

However, we can re-write the field variations in terms of
lattice variables. First, we define a continuum current
\begin{equation}
j_{\mu} = \gamma_{\mu} \vp \ .
\end{equation}
We now block this current by integrating the flux through
the face between two adjacent lattice cells,
\begin{equation} \label{curr}
J_{\mu ,x} = \int_{-1/2}^{1/2} j_{\mu}(x+ \frac{1}{2} \hat \mu
+u_{\nu}) \ du_{\nu} \qquad (\nu \neq \mu ,~ \vert \hat \mu \vert =1).
\end{equation}
This is a perfect lattice current \cite{Schwing}. Here we assume
it to be implemented 
so that eq. (\ref{curr}) holds exactly.
As an interesting property, 
its lattice divergence is equal to the blocked continuum divergence,
\begin{equation}
\nabla_{\mu} J_{x,\mu} := \sum_{\mu} (J_{\mu ,x}-J_{\mu , x-\hat \mu})
= \int_{C_{x}} \partial_{\mu} j_{\mu}
(u) \ du \qquad ({\rm Gauss'~law}).
\end{equation}
We are now prepared to write the variation of the lattice fermion
field from eq. (\ref{ptrafo}) in terms of lattice variables,
\begin{equation}
\delta \Psi = - \nabla_{\mu} J_{\mu} \ \eps \ , \quad
\delta \bar \Psi = - \epsb \ \nabla_{\mu} J_{\mu} \ .
\end{equation}
In addition, we introduce the fermionic lattice field
\begin{equation}
\tilde \Psi_{x} := \int_{C_{x}} \ \psi (u) du \ ,
\end{equation}
which allows us to write also $\delta \Phi$ in terms of lattice
quantities,
\begin{equation}
\delta \Phi = \epsb \ \tilde \Psi \ .
\end{equation}
For a $\delta$ function RGT in the fermionic sector we have
$\Psi = \tilde \Psi$, but for finite $\alpha^{f}$ this
does not hold exactly.\\

A generalization is possible for instance with respect to the blocking
scheme. Instead of the block average scheme we have used so far,
we can start from a more general ansatz
\begin{eqnarray} && \hspace{-20mm}
 T = \sum_{x,y} [ \bar \Psi_{x} - \int \Pi^{f}(x-u) \bar \psi (u) du ]
\ (\alpha^{f})^{-1}_{xy} \ [ \Psi_{y} - \int \Pi^{f}(x-u) \psi (u) du ] 
\nonumber \\ && \hspace{-15mm}
+ \sum_{x,y} [ \Phi_{x} - \int \Pi^{s}(x-u) \vp (u) du ]
\ (\alpha^{s})^{-1}_{xy} \ [ \Phi_{y} - \int \Pi^{s}(x-u) \vp (u) du ] \ ,
\end{eqnarray}
where $\int \Pi^{f}(u) du = \int \Pi^{s}(u) du =1$. Both
convolution functions $\Pi^{f}$, $\Pi^{s}$ are peaked around zero
and decay fast enough to preserve locality on the lattice.

Correspondingly, the variations of the lattice fields turn into
\begin{eqnarray}
\delta \Psi_{x} &=& - \gamma_{\mu} \int \Pi^{f}(x-u) \partial_{\mu}
\vp (u) du \ \eps \ , \nonumber \\
\delta \Phi_{x} &=& \epsb \int \Pi^{s}(x-u) \psi (u) du 
:= \epsb \ \tilde \Psi_{x} \ ,
\end{eqnarray}
where we have adjusted the definition of $\tilde \Psi$.
If we want to achieve $\Psi = \tilde \Psi$, then we need -- except for
$\alpha^{f} =0$ --
also $\Pi^{f}=\Pi^{s}$.

This generalized scheme does not yield an obvious
lattice current any more; the latter is a virtue of the block average
scheme (characterized by $\Pi^{f}(u)$, $\Pi^{s}(u) =1$ if $u\in
C_{0}$, and 0 otherwise).
In the general case, it is easier to consider the continuum
divergence
\begin{equation}
d(u) = \partial_{\mu} j_{\mu}(u) = \gamma_{\mu} \partial_{\mu} 
\vp (u) \ ,
\end{equation}
and build from it directly the lattice divergence
\begin{equation}
D_{x} = \int \Pi^{f}(x-u) d(u) \ du \ , 
\qquad \delta \Psi = -D \ \eps \ .
\end{equation}

Regarding the transformation algebra, we obtain
\begin{equation} \label{ptranslat}
[\delta_{1},\delta_{2}] \ \Phi_{x} = ( \bar \epsilon_{1} \ 
\gamma_{\mu} \ \epsilon_{2} \ - \ \bar \epsilon_{2} \ 
\gamma_{\mu} \ \epsilon_{1})
\int \Pi^{s}(x-u) \partial_{\mu} \vp (u) \ du \ .
\end{equation}
In particular, for the block average scheme this simplifies to
\begin{equation} \label{ptranslatba}
[ \delta_{1} , \delta_{2} ] \ \Phi
= \bar \epsilon_{1} \ \nabla_{\mu}
J_{\mu} \ \epsilon_{2} - \bar \epsilon_{2} \ \nabla_{\mu}
J_{\mu} \ \epsilon_{1} \ .
\end{equation}
We see that the continuum translation operator is inherited by
the perfect lattice formulation in a consistent way:
eqs. (\ref{ptranslat}), (\ref{ptranslatba}) 
show that the corresponding lattice translation operator
is just the blocked continuum translation operator.
The formula for $[\delta_{1},\delta_{2}] \Psi$ is analogous. 
If we require the resulting translation operators to be identical,
then we need $\Pi^{f} = \Pi^{s}$.
Then the algebra of field variations and translation closes
under the blocking integral.

In any case, the fermionic and scalar spectrum are equal, because
they both coincide with the continuum spectrum.\\

It is straightforward to apply this perfect
lattice formulation to more complicated cases, see below. Interactions
can be included perturbatively in the process of blocking
from the continuum. For asymptotically free theories,
at $m=0$ even the classically perfect action
behaves perfectly \cite{Has}. Hence by means of an implicit
(but not just symbolic) definition of the action -- in terms of
classical inverse blocking -- we can also go beyond perturbation 
theory in the massless case. This is analogous to the
fixed point formulation of a chiral gauge theory on the lattice
\cite{Schwing,HLN}.

\section{Adding an auxiliary scalar field}

To proceed to the 2d Wess-Zumino model, we include
an auxiliary scalar field.
\footnote{This is even necessary for the general validity
of the resulting translation operator; otherwise it only holds 
on-shell. See for instance P. Freud, ``Introduction to
Supersymmetry'', Cambridge University Press, 1986.}
This equilibrates the number of fermionic and bosonic degrees of
freedom.
The continuum Lagrangian and the field variations read
\begin{eqnarray}
{\cal L} &=& \bar \psi \gamma_{\mu} \partial_{\mu} \psi +
\partial_{\mu} \vp \partial_{\mu} \vp + f^{2} \ , \nonumber \\
\delta \psi &=& - (\gamma_{\mu}\partial_{\mu}\vp +f) \ \eps \ , 
\quad
\delta \vp = \epsb \ \psi \ , \quad \label{FF}
\delta f = \epsb \ \gamma_{\mu} \partial_{\mu} \psi \ .
\end{eqnarray}

Now the method presented in Ref. \cite{Jap} is applicable
in an extended form, if we use the following lattice discretization:
\begin{eqnarray}
S &=& \int_{B} \frac{d^{2}p}{(2\pi )^{2}} \ \Big[ \bar \Psi (-p)
i \gamma_{\mu} \bar p_{\mu} \Psi (p) + \Phi (-p) \bar p^{2} \Phi (p)
+ F(-p) F(p) \nonumber \\
&& + \bar \Psi (-p) W^{f}(p) \Psi (p)+2\Phi (-p) W^{s}(p) F(p)
- \Phi (-p) W^{s}(p)^{2} \Phi (p) \Big] \ , \nonumber \\
\delta \Psi (p) &=& -\{ [ i\gamma_{\mu}\bar p_{\mu} +W^{s}(p)]
\Phi (p) + F(p) \} \ \eps
\nonumber \\
\delta \Phi (p) &=& \epsb \ \Psi (p) \nonumber \\
\delta F(p) &=& \epsb \ [ i \gamma_{\mu} \bar p_{\mu} -W^{s}(p)]
\Psi (p) \ ,
\end{eqnarray}
where $W^{f}(p),\ W^{s}(p)$ are some sort of Wilson terms
(zero at the origin, non-zero at the edges of the Brillouin zone,
local and $2\pi$ periodic, which implies that they are even).
Hence they remove the degeneracy of the physical particles
with their doublers.
The standard form $1/2 \ \hat p^{2}$ is an example, but we can
also insert more general scalar and fermionic Wilson terms
\footnote{The continuum limit of this action does in fact correspond
to eq. (\ref{FF}), as we see if we substitute $F(p) \to \tilde F(p) =
F(p) + \Phi (p) W^{s} (p)$. Then the mixed term of $F$ and $\Phi$
disappears, and we obtain another irrelevant term 
$- \Phi (-p) W^{s}(p)^{2} \Phi (p)$.}
and we always arrive at $\delta S =0$. 
If we also want the fermion and scalar spectrum to coincide, 
then we have to relate $W^{f}$ and $W^{s}$.
The procedure applied in Ref. \cite{Jap} further restores a remnant
chiral symmetry by means of the so-called overlap formalism,
and this could also be done here.\\

Instead we can apply the perfect action machine from section 
\ref{perfect}, starting from the continuum system (\ref{FF}).
This also solves the doubling problem and maintains a (remnant)
chiral symmetry. The perfect propagator of the lattice field
$F$ reads 
\begin{equation} \label{statprop}
\Delta^{\bar s}(p) = \sum_{l \in \Z^{2}} 
\Pi^{\bar s}(p+2\pi l)^{2}+\alpha^{\bar s}(p) \ ,
\end{equation}
where $\alpha^{\bar s}$ is a RGT
parameter analogous to $\alpha^{s}$.
We now introduce {\em two} continuum currents, $\gamma_{\mu}\vp$ and
$\gamma_{\mu} \psi$, and we construct perfect currents 
from them. The explicit formulae are a straightforward extension
of the formulae in section \ref{perfect}. We do not display them here, but
we write them down for the further extension to the 4d Wess-Zumino model
in the appendix.


\section{Conclusions}

We illuminated the problem of a direct construction
of a supersymmetric lattice formulation.
Then we have shown how a construction in terms of
renormalization group transformations can be achieved.
We preserve invariance under a continuous supersymmetric
type of field transformations in a local perfect lattice action,
which has also a remnant chiral symmetry.
This applies to the 2d models discussed above, as well as to
the 4d Wess-Zumino model, see appendix. 
We remark that the perfect formulation also cures
the well-known problems related to the Leibniz rule 
\cite{susylat} 
\footnote{See in particular the first paper by S. Nojiri.} --
which breaks down for usual lattice discretizations --
because here we keep track of the exact continuum differential
operators.
This is manifest in the translation operator, which results from a
commutator of field variations: in the perfect lattice
formulation, we obtain
the consistently blocked continuum translation operator.
Therefore the algebra with the field variations closes.

Moreover, in the perfect lattice formulation, the fermionic and
scalar dispersion relation coincide automatically.

The next step is
the inclusion of the gauge interaction; this work is in progress.
A consistent blocking of the gauge fields from the continuum
leads to a perfect action with all the continuum symmetry properties
in the observables -- and also in the action, if the transformation
term respects these symmetries -- but this construction can only
be performed perturbatively. For asymptotically
free theories in the massless case, a classically
perfect action -- constructed by simplified inverse blocking 
(based on minimization) --
is sufficient for the same purpose and enables the step beyond
perturbation theory.\\

{\em Acknowledgment} \ I am indebted to M. Peardon for many helpful
discussions. I also thank him and I. Montvay for reading 
the manuscript.

\appendix

\section{Application to the 4d Wess-Zumino model}

For completeness, we write down the corresponding formulae
for a perfect lattice formulation of the 4d Wess-Zumino model:
\begin{eqnarray} \nonumber
&& \hsp
e^{-S[ \Psi ,\Phi^{(1)},\Phi^{(2)} ,F^{(1)},F^{(2)}]} = \int D \psi 
D \vp^{(1)} D \vp^{(2)} D f^{(1)} D f^{(2)} \
e^{-s[ \psi , \vp^{(1)}, \vp^{(2)} ,f^{(1)},f^{(2)}]} \times \nonumber \\ 
&& \hsp \exp \{ -T [ \Psi , \psi , \Phi^{(1)},\vp^{(1)},\Phi^{(2)},\vp^{(2)},
F^{(1)},f^{(1)}, F^{(2)},f^{(2)}] \} \ , \\
&& \hsp s[\psi ,\vp^{(1)},\vp^{(2)},f^{(1)},f^{(2)}] =
\frac{1}{2} \int \!\! d^{4}x \ [ \bar \psi \gamma_{\mu} \partial_{\mu}
\psi + ( \partial_{\mu} \vp^{(1)})^{2} + ( \partial_{\mu} \vp^{(2)})^{2}
+ f^{(1)~2} + f^{(2)~2} ] \ , \nonumber \\
&& \hsp T = \sum_{x,y} [ \bar \Psi_{x} - \int_{C_{x}} \bar \psi (u) du ]
\ (\alpha^{f})^{-1}_{xy} \ [ \Psi_{y} - \int_{C_{y}} \psi (u) du ] 
\nonumber \\
&& + \sum_{i=1}^{2} \sum_{x,y} \Big\{ [ \Phi^{(i)}_{x} - 
\int_{C_{x}} \vp^{(i)} (u) du ]
\ (\alpha_{i}^{s})^{-1}_{xy} \ [ \Phi^{(i)}_{y} - \int_{C_{y}} 
\vp^{(i)} (u) du ] \nonumber \\
&& \hspace{15mm} + [ F^{(i)}_{x} - \int_{C_{x}} f^{(i)} (u) du ]
\ (\alpha_{i}^{\bar s})^{-1}_{xy} \ [ F^{(i)}_{y} - 
\int_{C_{y}} f^{(i)} (u) du ] \ \Big\} \ ,
\end{eqnarray}
where  $\alpha_{i}^{s}$ and $\alpha_{i}^{\bar s}$ have to be positive,
and we are using the block average scheme.
The perfect action $S$ can be assembled by analogy from eqs.
(\ref{perfact}) and (\ref{statprop}).
The transformation of the continuum fields have the usual form,
\begin{eqnarray}
\delta \psi &=& - [\gamma_{\mu} (\partial_{\mu} \vp^{(1)} + \gamma_{5}
\partial_{\mu} \vp^{(2)}) + f^{(1)}+\gamma_{5} f^{(2)}] \
\eps \ , \nonumber \\
\delta \vp^{(1)} &=& \epsb \ \psi \ , \quad 
\delta \vp^{(2)} = \epsb \ \gamma_{5} \psi \ ,\nonumber \\
\delta f^{(1)} &=& \epsb \ \gamma_{\mu} \partial_{\mu} \psi \ , \quad 
\delta \vp^{(2)} = \epsb \ \gamma_{5} \gamma_{\mu} \partial_{\mu} \psi
\ ,
\end{eqnarray}
and the lattice field transformations amount to
\begin{eqnarray}
\delta \Psi_{x} &=& - \int_{C_{x}} \Big\{ \gamma_{\mu} [ \partial_{\mu} 
\vp^{(1)} (u) + \gamma_{5} \partial_{\mu} \vp^{(2)} (u)]
+ f^{(1)}(u) + \gamma_{5} f^{(2)}(u) \Big\} du \ \eps \nonumber \\
&:=& - \Big[ \nabla_{\mu} [J_{\mu ,x}+ J^{5}_{\mu ,x}]
+ \tilde F^{(1)}_{x} + \gamma_{5} \tilde F^{(2)}_{x} \Big] \ \eps 
\ , \nonumber \\
\delta \Phi^{(1)}_{x} &=& \epsb \int_{C_{x}} \psi (u) du
= \epsb \ \tilde \Psi_{x} \ , \quad
\delta \Phi^{(2)}_{x} = \epsb \ \gamma_{5} \int_{C_{x}} \psi (u) du
= \epsb \ \gamma_{5} \tilde \Psi_{x} \ ,
\nonumber \\
\delta F^{(1)}_{x} &=& \epsb \ \gamma_{\mu} \int_{C_{x}} \partial_{\mu}
\psi (u) du := \epsb \ \nabla_{\mu} I_{\mu ,x} , \nonumber \\
\delta F^{(2)}_{x} &=& \epsb \ \gamma_{5} \gamma_{\mu} \int_{C_{x}} 
\partial_{\mu} \psi (u) du = \epsb \ \gamma_{5} \nabla_{\mu} 
I_{\mu ,x} \ ,
\end{eqnarray}
where $J_{\mu},\ J^{5}_{\mu}$ and $I_{\mu}$
are perfect lattice currents.

Again the resulting translation operator is the blocked
continuum translation operators, it forms a closed algebra with
the field variations, the Leibniz rule is satisfied,
and the dispersion relations of $\Psi$ and $\Phi$ coincide.

\end{document}